\documentclass[pdflatex,sn-mathphys-num]{sn-jnl}

\usepackage{graphicx}%
\usepackage{multirow}%
\usepackage{amsmath,amssymb,amsfonts}%
\usepackage{amsthm}%
\usepackage{mathrsfs}%
\usepackage[title]{appendix}%
\usepackage{xcolor}%
\usepackage{textcomp}%
\usepackage{manyfoot}%
\usepackage{booktabs}%
\usepackage{algorithm}%
\usepackage{algorithmicx}%
\usepackage{algpseudocode}%
\usepackage{listings}%
\usepackage[title]{appendix}%

\begin{document}

\title{In-Network Memory Access: Bridging SmartNIC and Host Memory}

\author[1]{\sur{Mohammed Zain Farooqi}}

\author[2]{\sur{Masoud Hemmatpour}}

\author[3]{\sur{Tore Heide Larsen}}

\affil[1]{\orgdiv{Department of Physics}, \orgname{University of Oslo (UiO)}}

\affil[2]{\orgdiv{Department of Computer Science}, \orgname{Arctic University of Norway (UiT)}}

\affil[3]{\orgdiv{HPC Department}, \orgname{Simula Research Laboratory (SRL)}}


\abstract{SmartNICs have been increasingly utilized across various applications to offload specific computational tasks, thereby enhancing overall system performance. However, this offloading process introduces several communication challenges that must be addressed for effective integration. A key challenge lies in establishing efficient communication between the offloaded components and the main application running on the host. In this study, we evaluate different approaches for achieving memory access between the host and SmartNIC. We analyze memory access performance on both the SmartNIC and the host to support in-network applications and guide the selection of an appropriate memory access design.}

\keywords{In-network computing, FPGA, SmartNIC, DPU}



\maketitle

\section{Introduction}
Traditional data networks passively transport data from one node to another, and the role of computation within such networks is extremely limited \cite{activenetwork}. The discussion on computation in the network devices has existed since 1996 through the active networks architecture
and Open Signalling communities  \cite{prognet,activenetwork,opensig}. Active Networks allowed
The network to perform customized computations on the user data \cite{activenetwork}. Open Signalling was a series of workshops with the goal of making ATM, Internet, and mobile networks more open, extensible, and programmable \cite{opensig}. Although such discussion existed for a long time, the real hardware to run computation in the network devices were not available. Programmable Network Devices (PNDs), such as SmartNICs and programmable switches, have recently enabled the execution of customized computations directly on network hardware. This capability is driving the emergence of In-Network Computing (INC)—also known as In-Network Computation or NetCompute—which refers to running user-defined programs within these PNDs \cite{inc}.

There are several studies that offload processing to SmartNIC. Table \ref{tab:smartnic} summarizes some of these works. Offloading offers several advantages, including reducing the load on the host CPU, bringing computation closer to the network, decreasing traffic, and enhancing security. However, the offloaded tasks on the SmartNIC still need to maintain communication with the host. The communication between the host and SmartNIC is especially crucial when it comes to memory access. SmartNICs and hosts frequently need to access each other’s memory to process data efficiently. This interaction requires low-latency, high-bandwidth memory access to avoid bottlenecks that could degrade overall system performance. In scenarios where SmartNICs handle significant amounts of data, effective memory access mechanisms are key to maintaining high throughput, minimizing latency, and ensuring smooth operations. Various studies employ different approaches for communication depending on the SmartNIC: In \cite{bayatpour2021bluesmpi},  they utilize Remote Direct Memory Access (RDMA) communication for accessing SmartNIC memory, while Lake and ipipe use Direct Memory Access (DMA) to access host memory \cite{lake,liu2019ipipe}. \cite{mandalstorage} exploits Virtual Function I/O (VFIO) for communication.

While prior work has focused on evaluating the standalone performance of SmartNICs \cite{liu2021performance,xu2024performance}, little attention has been given to the communication performance between the host and SmartNIC. To the best of our knowledge, this paper presents the first comprehensive analysis of various memory access techniques between the host and SmartNIC. This analysis offers important insights into the communication dynamics between host and SmartNIC. We focus on two important type of SmartNICs and run experiments based on the characteristics of the real workload in datacenters. Our main contributions are as follows:

\begin{itemize}
    \item Exploring potential communication methods between the host and SmartNICs
    \item Creating scenarios to assess host/SmartNIC memory access 
    \item Conducting evaluation and analysis of different scenarios
\end{itemize}

The remainder of this paper is structured as follows: Section \ref{sec:smartnic} provides an overview of SmartNICs, particularly FPGA-based and SoC-based SmartNICs. Section \ref{sec:back} describes background and relevant requirements for this study. Section \ref{sec:comtype} outlines the possible techniques for memory access between the Host and the SmartNIC. Section \ref{sec:exp} describes experiment design in FPGA and SoC-based SmartNICs. Section \ref{sec:eval} presents the evaluation of different experiments. Finally, in Section \ref{sec:conc} we draw conclusions.

\begin{table*}[h]
 \centering
  \renewcommand\arraystretch{1.1}
		\centering
\begin{tabular}{ |l|l|l|l| }
\hline
\centering
\textbf{Name} & \textbf{Use case} & \textbf{Target} & \textbf{Refs}  \\ \hline

INCS & bot traffic detection & BlueField &  \cite{hemmatpour2024commerce} \\
BluesMPI & MPI communication  & BlueField &  \cite{bayatpour2021bluesmpi} \\
Lake & caching & NetFPGA-SUME & \cite{lake}  \\
KV-DIRECT & caching &  Intel Stratix V FPGA &  \cite{kvdirect}\\ 
Xenic &  communication/processing  & LiquidIO 3 & \cite{schuh2021xenic} \\
NICA & inline processing & Mellanox Innova with Xilinx FPGA  & \cite{eran2019nica} \\
iPipe &  distributed applications & LiquidIOII CN2350/CN2360 &  \cite{liu2019offloading}\\
OXDP &  packet processing  & Netronome Agilio CX 40GbE & \cite{wang2023oxdp}\\
LineFS &  file systems  & BlueField & \cite{kim2021linefs}\\
UNO  &  networking functions &  Mellanox TILEGx36 & \cite{le2017uno}\\ 
Clara  &  insights for NF  & SOC-based SmartNICs  & \cite{qiu2021automated}\\
\hline
\end{tabular}
\vspace{-5pt}
\caption{Task offloading enabled by SmartNICs.}
\label{tab:smartnic}
\end{table*}

\section{SmartNIC} \label{sec:smartnic}

The term SmartNIC implicates that it is an extension of standard Network Interface Card (NIC) devices with intelligence or the ability to act somewhat smart to solve complex tasks \cite{doring2021smartnics,katsikas2021you}. This was achieved by adding a processing
unit to a NIC. SmartNICs have emerged as a revolutionary technology to enhance the computational capabilities of NIC by offloading certain processing tasks \cite{kfoury2024comprehensive}. Vendors have their own definition and categories of SmartNIC based on their processing unit \cite{industrysmartnic}. The terms accelerators and engines
are also used to refer to domain-specific processors \cite{ibanez2019case}. In the following, the various processing units utilized in SmartNICs are described.

\begin{itemize}
    \item \textbf{Field Programmable Gate Arrays (FPGAs)} are suitable to process data flows at high speed. FPGA-based SmartNICs have been used in different research studies \cite{lant2019toward,lake} and industry \cite{microsoftfpga,caulfield2018beyond}. The most of existing SmartNICs are equipped with Xilinx (AMD) \cite{xilinxfpga} and Intel \cite{intelfpga} FPGAs and are programmed with Hardware Description Languages (HDL) \cite{palnitkar2003verilog}.

    \item \textbf{General-purpose CPU} process data in a traditional manner through high-level languages such as C/python. These are suitable to offload complicated tasks from the host to the CPU \cite{bayatpour2021bluesmpi}.

    \item \textbf{ Application-Specific Integrated Circuit (ASICs)}
    offer specialized and highly efficient processing capabilities. However, its flexibility is limited to predefined functions. There are not many ASIC-based smart network cards. Mellanox’s ConnectX series, Broadcom’s NetXtreme series, and Cavium’s FastLinQ series are important ASIC vendors in SmartNICs. 
    \item \textbf{Graphics Processing Unit (GPU)} has become a key element in, e.g., Artificial Intelligence (AI) in a way GPU-oriented SmartNICs have been designed to accelerate
applications running on distributed GPUs \cite{wang2022fpganic}. Furthermore, SmartNICs, such as BlueField, have been developed with integrated GPUs as processing units to enhance computational capabilities.

    \item \textbf{Domain Specific Processor (DSP)}
  include regular expression (RegEx) used for tasks requiring Deep Packet Inspection (DPI), Non-Volatile Memory Host Controller over Fabrics (NVMe-oF) for remote storage, data compression, data deduplication, Remote Direct Memory Access (RDMA), etc.

      \item \textbf{Data Processing Unit (DPU)}
  is designed to enhance data center performance by offloading network, storage, and security workloads from the CPU. There are a few DPU-based SmartNICs in the market, for example, AMD Pollara 400, which exploits AMD Pensando DPUs. Pensando is capable of running P4 programs natively in their Match Processing Units (MPUs) \cite{pensando}.
\end{itemize}

In this paper, we focus primarily on FPGA-based SmartNICs while also discussing SoC-based variants. This is because FPGAs provide a distinctive blend of high performance, programmability, and flexibility, making them well-suited for advanced applications. Meanwhile, general-purpose CPUs (referred to as System on a Chip (SoC) in this paper) are also widely used due to their ease of programming. FPGA-based  SmartNICs are widely adopted in both research and industry, with notable implementations such as Microsoft’s deployment \cite{lake,micro,putnamretrospective}.

\subsection{FPGA-based SmartNIC} \label{sec:fpgasmartnic}

\begin{figure}[h]
  \centering
\includegraphics[scale=0.5]{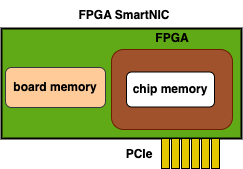} 
\vspace{-0.5em}
  \caption{Available memories on FPGA-based SmartNICs.}
  \label{fig:fpgaarc}
\vspace{-3pt}
\end{figure}

FPGA-based SmartNICs bring a configurable and programmable hardware approach to datacenter architecture. FPGAs are available with or without a Microprocessor embedded on their fabric for on-chip computations alongside their programmable logic, which allows huge tasks to run in parallel. These make FPGA-based SmartNICs a good choice for offloading massive computations from host systems. Soft CPUs can be implemented on FPGAs even in the absence of embedded hard processors, allowing the design of systems comparable to SoC counterparts while still taking advantage of the FPGA’s configurable logic components. There are different on-chip and on-board memories in an FPGA-based SmartNIC, as Figure \ref{fig:fpgaarc} shows. The amount of on-chip memory depends on the resources available on the FPGA fabric at any moment. Such on-chip memory can be used as fast cache memory that can be configured for data processing with extremely low latency. FPGA-based SmartNICs also come with huge amounts of on-board (off-chip) DRAM that can be accessed in parallel by configuring concurrent memory controllers on the FPGA.

SmartNICs are connected to the host system via PCIe lanes, using x8 or x16 configurations, or x32 lanes in recent PCIe configurations. These PCIe lanes enable high-speed, low-latency data transfer between the SmartNIC and the host. Network data is received through the physical interface and processed on the FPGA and then transmitted to the host via PCIe. A variety of SmartNIC products are available on the market, including the NVIDIA Innova-2 Flex \cite{innova} and AMD Alveo \cite{amd_alveo_u250}.

\subsection{SOC-based SmartNIC} \label{sec:socsmartnic}
\begin{figure}[h]
  \centering
\includegraphics[scale=0.5]{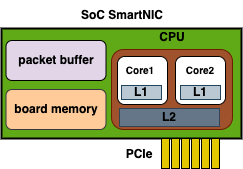} 
\vspace{-0.5em}
  \caption{Available memories on SOC-based SmartNICs.}
  \label{fig:socarc}
\vspace{-3pt}
\end{figure}

SoC-based SmartNICs integrate general-purpose processors, offering ease of programmability and compatibility with existing software stacks, making them ideal for the flexible and rapid development of network functions. As Figure \ref{fig:socarc} shows, generally there are three categories of memories on the board: (1) local and shared cache memories in the computing resource, which has a limited size but
access speed. This memory is accessible through the processing unit and can not be directly accessed from the host. (2) Packet buffer, which is onboard SRAM, along with fast indexing. However, it does not exist in all SmartNICs; depends on the vendor. (3) local DRAM, which is accessed via the onboard high-bandwidth coherent memory bus.

There are various SmartNIC products in the market, including Marvell (Cavium)’s LiquidIO \cite{marvel}, NVIDIA’s Bluefield-X \cite{nvidia}, Netronome’s Agilio \cite{netronome}, Huawei’s IN5500 \cite{Huawei}, and Broadcom’s Stingray \cite{broadcom}.

\section{Background} \label{sec:back}
In this section, we present the key concepts and fundamentals required to comprehend the topics discussed in this paper. We will outline the core principles associated with in-network memory access.

\subsection{AXI}

For data transfer between on-chip devices, high-speed, low-latency buses and communication protocols are used. Within a chip, the microprocessor interacts with peripherals through memory-mapped input/output (MMIO), where peripherals are mapped into the system's memory address space. In this context, memory-mapped means that the master device specifies an address within each transaction \cite{crockett2014zynq}. A commonly used interface for this purpose in FPGA designs is AXI. An interface protocol defined by ARM, AXI stands for Advanced eXtensible Interface, which is a part of the AMBA  standard (Advanced Microcontroller Bus Architecture). AXI is a high-performance, high-frequency bus protocol and not the bus or the interconnect itself,  for on-chip communications between managing and subordinating peripherals in system designs. Separate read and write channels, and multiple addresses to connected slaves, improve system performance and bandwidths as it allows parallel processing of transactions, by decoupling the reads and writes into separate channels. The Advanced Microcontroller Bus Architecture (AMBA) AXI specification provided by ARM \cite{arm_axi} outlines the key features of the protocol:

\begin{itemize}
    \item suitable for high-bandwidth and low-latency designs,
    \item high-frequency operation without using complex bridges,
    \item support for interfacing a wide range of components,
    \item lower latency access to memory controllers.
\end{itemize}

Depending on how much data is being transferred between the communicating memory mapped, devices we have different types of AXI interfaces.

\begin{itemize}
    \item \textbf{AXI4-Lite}: for low throughput, simple register memory-mapped communication. One-word transfers or single-beat transfers.
    \item \textbf{AXI4}: for burst transfers (multiple beats, data words), high-performance memory-mapped transfers. 
    \item \textbf{AXI4-Stream}: for high-speed point-to-point streaming data.
    \item \textbf{AXI5}: extends prior specification generations and add several important performance and scalability features
    \end{itemize}

In a typical FPGA system design, various Intellectual Property (IP) cores are connected using these protocols. For instance, serial interface peripherals and timers communicate with the CPU via AXI4-Lite, while memory controllers and DMA controllers use AXI4 for interfacing within the same system.

\subsection{PCIe}
Peripheral Component Interconnect Express (PCIe) is a high-speed bus expansion standard developed by Intel \cite{harris2010digital} for connecting off-chip peripheral devices—such as sound cards, memory modules, Ethernet NICs, etc. The PCIe standard replaces older bus architectures such as PCI, PCI-X, and AGP, which were based on parallel communication using fixed copper slots where multiple wires transmitted signals simultaneously \cite{jackson2012pci}. In contrast, PCIe uses a serial interface with multiple full-duplex lanes, offering significantly higher throughput (e.g., PCIe Gen3 at 16 GB/s vs. PCI at 133 MB/s \cite{harris2010digital}), better clocking compatibility, and enhanced signal synchronization compared to the parallel architecture of legacy PCI buses.

Devices connected over PCIe communicate via an interconnect called a link or serial lane, which is made up of one or more duplex i.e., transmit and receive pairs of connections. Each link consists of only four wires: two for receiving data and two for transmitting data. This small number contrasts with an earlier version of PCI that consisted of 64 wires \cite{Patterson2020}, which was called a parallel bus. These lanes are made up of 1, 2, 4, 8, 16, or 32 in number represented as x1, x2, x4, x8, x16, and x32, called the link width of the PCIe interface. PCIe devices communicate with the CPU over these lanes  in the form of serialized, packed transmissions via switches and root complexes, which do the intermediary steps of the PCIe link transfer initializations and configurations. These lanes provide a speed of 1 GB/s individually, aggregating the total bandwidth according to their link width and PCIe generation revision. PCIe packets called Transaction Layer Packets or TLPs carry information, for example, I/O read or write, memory read/write/completion, interrupt, etc. Each TLP has a header of about 3-4 DWs (Data Words, 32 bits) and a payload of up to a maximum of 4096 bytes. The header contains information about the type of TLP operation and whether it contains data or not. It also contains the requester ID of the devices that requested data to be read from the memory location being communicated over PCIe.

PCIe is a three-layer protocol \cite{pci_express_tech_3_0}, operating across the Transaction Layer, Data Link Layer, and Physical Layer of the OSI model, which together constitute the PCIe protocol stack.

\begin{itemize}
    \item \textbf{Transaction Layer}: This layer handles the generation and processing of PCIe requests, such as memory reads/writes, and ensures proper data delivery between the connected devices. It manages the creation of PCIe packets and routes them appropriately based on the destination.
    \item \textbf{Data Link Layer}: This layer provides reliable communication between PCIe devices by establishing and maintaining a connection. It ensures data integrity through error detection and acknowledgment mechanisms to make sure successful packet transmissions happen.
    \item \textbf{Physical Layer}: This layer is responsible for the actual transmission of data across PCIe lanes using serial communication. It converts packets from the data link layer into electrical signals and handles the low-level signaling, lane negotiation, and clocking mechanisms over differential pairs of the lanes.
\end{itemize}

 The two layers of the physical and link layers are implemented in the PCIe IP core in FPGAs, allowing seamless PCIe communication between the FPGA and other devices. The IP core integrates hardware circuitry and logic, managing low-level functions such as data transmission, error correction, and lane negotiation. The transaction layer, which handles the generation, reception, and routing of PCIe packets, is also part of the FPGA's hardware implementation alongside interfacing with user applications through custom logic, enabling efficient data handling and communication with software. 

Such PCIe IP core in FPGAs is often implemented alongside Direct Memory Access (DMA) cores to enhance data transfer efficiency between the FPGA and the host system. By integrating DMA capabilities, these cores enable the FPGA endpoints, such as FPGA-based SmartNICs, to transfer large volumes of data directly to and from host memory without CPU intervention, thus reducing latency and freeing up processing resources for other tasks. FPGA-based SmartNICs allow us to design multiple hardware channel DMA IP cores that allow aggregated PCIe transactions over the lanes, significantly increasing bandwidth. Such offloading of memory transfers to DMA cores allows the CPU to efficiently transfer data to and from the SmartNIC, helping us achieve offloading of data and applications from the host.

\subsection{DMA}
Direct Memory Access (DMA) is the capability of CPU bus architecture that allows data to be sent directly from one memory location or peripheral to another memory location for example from the local memory of the processor to external memory or hard disk drive or any other peripheral that is connected to the host, traditionally, this was the responsibility of the CPU to handle transfers between the peripheral devices that would create a bottleneck affecting the whole system,  but by leveraging DMA controllers, we can now have peripherals transfer data between them without involving the CPU.

In this process, DMA acts like traffic control to handle data between devices, and the CPU is freed from involvement with the data transfer, allowing the peripheral memories to communicate with each other using the DMA controller, speeding up the computer operations, which would have been hindered due to the memory transactions. This is possible via DMA controllers and hardware peripherals that access the computer bus for the data transfers. DMA devices are partially dependent on co-processors which offload data transfers from the main CPU, offering high-performance capabilities \cite{noergaard2012embedded}. These controllers are sometimes incorporated into the master processor or implemented as separate devices.

Common types of transfers performed by a DMA controller include:
\begin{itemize}
    \item Internal to external memory to memory transfers,
    \item I/O devices to memory or memory to I/Os,
    \item Transfers between communication ports to memory, for example, serial ports and memory.
\end{itemize}

The initial setup of DMA operations makes the CPU write to DMA registers with setup information, including starting address, data to transfer, type of transfer, mode of transfers, the direction of the transfers, and destination addresses \cite{noergaard2012embedded}.

BUS master DMA is a type of DMA often deployed by processor architectures for memory access between peripherals, where memory is shared over internal data buses or external high-speed interfaces to connected devices. For example, in PCIe bus  DMA transactions, the DMA controller goes through the following steps:

\begin{itemize}
    \item The DMA controller requests the bus when it starts a transfer as a master device.
    \item The host CPU completes any current transfers and grants the bus to the DMA controller.
    \item The DMA transfers (i.e., host to card and card to host)  data words until the transfers last or the CPU requests the bus for its own transactions. If the bus is granted back to the CPU, it is requested again to complete the transfers. Once the bus is granted, the process of memory access to the CPU isn't involved anymore.
    \item DMA controllers notify the CPU about finishing transfers with interrupts to the CPU, and the system bus is released.
\end{itemize}

The most common mode of DMA used in memory-to-memory transfers between host and card is scatter-gather mode, which works with the DMA controller acting as a master on the PCIe endpoint hardware. The necessary drivers supporting the DMA controller specify DMA transfer parameters, Source (SRC) and Destination (DST) descriptors. The SRC and DST descriptors are programmable registers that inform the DMA controller about the data size, buffer locations, and the source and destination addresses for reading from and writing to the connected PCIe endpoint. The DMA channels use these descriptors to initiate data transfers from the source addresses to the corresponding destination addresses. Since these descriptors can reside at non-contiguous memory locations—depending on memory availability—the method is referred to as scatter/gather DMA. A scatter-gather based Bus Master DMA (BMD) is the most common type of DMA found in systems based on PCIe \cite{amd_pcie_versal_dma}. 

\subsection{RDMA}
High-speed data transfer is hindered by TCP/IP protocols because of their high CPU and memory consumption. This constraint emerged a new category of network fabrics, using a technology known as Remote Direct Memory Access (RDMA). RDMA allows data to be transferred directly between the memory of two computers or devices without involving their CPUs, caches, or operating systems \cite{hemmatpour2020analyzing}. Examples of such network fabrics are internet Wide-Area RDMA Protocol (iWARP) \cite{iwarp}, RDMA over Converged Ethernet (RoCE) \cite{roce}, and InfiniBand \cite{inf}. In contrast with conventional protocols, RDMA implements the entire transport logic in network interface card—commonly known as Host Channel Adapter (HCA)—to boost the performance.

Generations of RDMA technology, primarily through InfiniBand, have progressively increased data transfer speeds. Starting with SDR (Single Data Rate) at 10 Gbps, speeds doubled with DDR (Double Data Rate) to 20 Gbps, and then QDR (Quad Data Rate) reached 40 Gbps. FDR (Fourteen Data Rate) improved encoding efficiency, offering 56 Gbps, followed by EDR (Enhanced Data Rate) at 100 Gbps. HDR (High Data Rate) doubled that to 200 Gbps, and the latest, NDR (Next Data Rate), achieves 400 Gbps \cite{inf}. Each generation offers higher bandwidth, lower latency, and improved efficiency for high-performance computing and data center applications.

\section{Host and SmartNIC Memory Access}\label{sec:comtype}
This section explores the memory access and communication mechanisms between the host and SmartNICs, with a particular focus on FPGA-based and SOC-based implementations. In FPGA-based implementations, we primarily center our analysis on AMD Xilinx platforms, as they are widely adopted across various applications. We will explore how these architectures facilitate efficient data transfer and processing, highlighting the differences and advantages of each approach.

\subsection{FPGA-based SmartNIC}
A variety of FPGA-based SmartNICs are available, incorporating Intel or AMD (formerly Xilinx) FPGAs, including products such as Napatech SmartNICs, AMD Alveo, and NVIDIA Innova Flex. AMD Xilinx has IPs (i.e., XDMA and QDMA) based on bus master DMA whose architectures are built around the PCIe IP cores. These allow DMA between the host and FPGA memory over the PCIe interconnect.

\subsubsection{XDMA}
The AMD Xilinx DMA/Bridge Subsystem for PCI Express (XDMA) is a high-performance, channel-based DMA IP that integrates a scatter-gather DMA engine with the PCI Express block. It enables efficient communication and data transfer between FPGAs and host systems. XDMA IP supports AXI4 memory-mapped and AXI streaming interfaces, enabling flexible communication between PCIe and AXI memory spaces. It is configurable as either a high-performance Direct Memory Access (DMA) data mover or a bridge between PCIe and AXI memory \cite{amd_pcie_dma}.

The XDMA subsystem is built on top of the PCIe interface, providing high-speed connectivity between the FPGA and external systems. It supports up to four Host-to-Card (H2C) and Card-to-Host (C2H) channels, enabling concurrent packet transfers over PCIe. Each channel appears as a separate device on the host side, allowing it to initiate and manage transfers independently. By interleaving channels across PCIe lanes, the XDMA IP maximizes bandwidth utilization by concurrently processing Transaction Layer Packets (TLPs) from multiple channels, effectively saturating the PCIe bus. 

The XDMA IP supports both AXI4 memory-mapped and AXI streaming interfaces for efficient access to FPGA memory peripherals, including DDR, BRAM, URAM, and other memory-mapped resources. The AXI4 memory-mapped interface facilitates direct read and write operations, ensuring high-speed transactions and data consistency across multiple channels. Conversely, the AXI streaming interface enables continuous, low-latency data flows, ideal for real-time processing applications that require high-bandwidth throughput. 

By leveraging PCIe’s multi-lane architecture, XDMA achieves high data throughput while maintaining low latency, making it a powerful data mover for high-performance FPGA-based systems. Its integration with AXI interfaces allows seamless communication between host systems and memory-mapped peripherals within the FPGA, supporting advanced data processing workflows and enabling efficient memory access in complex applications.

\subsubsection{QDMA}

Queue based Direct Memory Access subsystem (QDMA) is a PCIe-based DMA engine that is optimized for high-bandwidth and high packet count data transfers. As discussed earlier, the mechanism of XDMA uses the descriptors (instructions) provided by the host operating system to transfer data in either direction to and from the device. The QDMA uses queues, derived from queue set concepts of RDMA\cite{amd_qdma}. The DMA descriptors used for transfers are loaded with the queues, which are assigned as resources to PCIe Physical Functions (PFs) and Virtual Functions (VFs) by the IP. The subsystem is used with an AMD-provided reference driver. It supports both  AXI4 and AXI4-Lite interfaces.

The QDMA IP maps external memory into register-mapped space, allowing the FPGA to access memory as if it were a register array, using Base Address Registers (BARS). These BARS are a set of registers in the IP that hold addresses of memory regions that the QDMA  IP needs to access. The QDMA IP uses the AXI memory-mapped interface AXI-MM to map the external memory or any instantiated memory, into a register-mapped space. The QDMA drivers on the host pick recognize these BARS from the design when it is correctly configured, allowing us to do DMA transactions over PCIe to access AXI-MM peripherals from the host system.

QDMA differs from XDMA primarily in its queue based architecture, which is designed to support higher packet counts and manage complex data flows more efficiently. While XDMA uses a traditional descriptor based approach to manage data transfers, QDMA implements a queue based system that allows dynamic handling of multiple data streams. This queue based system is derived from RDMA principles and enables QDMA to manage data transfers through dedicated queues, which can be assigned to both PFs and VFs. While XDMA’s simpler, descriptor based model is typically used for straightforward data transfers using Physical channels. In both IPs, the use of AXI4 and AXI4-Lite interfaces allows for compatibility with AXI peripherals.

\subsection{SOC-based SmartNIC}
The type of data transfer interface offered by a SoC-based SmartNIC can vary, with some supporting RDMA verbs, native DMA primitives, or a combination of both. For instance, BlueField and Stingray primarily expose RDMA verbs instead of native DMA interfaces \cite{liu2019ipipe}. We focus primarily on NVIDIA's BlueField solution, as it is widely adopted in both research and industry applications. BlueField has recently introduced DMA support, enabling direct data transfers between the host and the SmartNIC \cite{dmadoca}.
                         
\subsubsection{RDMA}
SmartNIC and host memory can be accessed directly through RDMA operations. RDMA allows different operations on remote memory, such as read, write, and atomic operations (i.e., Fetch And Add (FAA) and Compare And Swap (CAS)). The receiving side’s CPU is completely passive during the memory access. Write operation to remote memory has no notification on the remote side. However, sending immediate data with a write operation notifies the remote side about the operation. The write and immediate notification are an atomic unit; either both arrive or neither arrives. Read operation reads a contiguous block of memory (in virtual address space) from the remote side and writes it to the specified local memory buffer.
\subsubsection{DMA}
BlueField supports DMA operations through the Data-Center-on-a-Chip Architecture (DOCA) framework. DOCA DMA enables efficient, low-latency, high-bandwidth data transfers between the host and the SmartNIC.  DOCA DMA enables a range of memory operations, including memory-to-memory copy, remote memory access, and scatter/gather transfers, all with minimal CPU involvement. This capability is particularly valuable in data-intensive applications such as storage and networking. By abstracting complex low-level operations, DOCA DMA allows developers to build high-performance, zero-copy data pipelines that leverage BlueField’s embedded processing and acceleration features.

\section{Experimental Design}\label{sec:exp}
In this section, we present the design of the experiments aimed at evaluating memory access mechanisms in both FPGA-based and SoC-based SmartNICs. Our goal is to analyze and compare the performance characteristics of different approaches in terms of data transfer. By implementing and testing representative scenarios on both platforms, we aim to gain deeper insights into their strengths and trade-offs for in-network computing applications.

\subsection{FPGA-based SmartNIC}
In our experiments, we utilize the AMD Xilinx Alveo U250 \cite{amd_alveo_u250}. It is a custom-built Ultrascale+ FPGA card housing the XCU250 FPGA-based on 16nm UltraScale™ architecture. We utilize the different on-chip and off-chip memories on the  Xilinx FPGA-based SmartNIC to evaluate memory access between the host and FPGA  memories.  The Alveo card is equipped with a Gen 3 PCIe interface, 4 DDR4 DIMMS, flash memories, and QSFP connector hardware for high-speed Ethernet connectivity \cite{amd_u200_u250}. 

The experiments aim to utilize the Alveo card’s BRAM, DDR4 DRAM, and their access from host systems utilizing DMA IPs in different case scenarios where the IPs are used as stand-alone systems or with additional peripherals for on chip data processing. Our methodology will include designing systems and DMA flows to effectively utilize each memory type and explore configurations where memory access is managed through both direct FPGA logic and microcontroller (or embedded processor) units, as well as configurations running with PetaLinux for more computation power while benchmarking the bandwidth performance of data transfers.

For this work, we adopt a low-level, board-aware design flow using Xilinx Vivado \cite{VivadoUG904}, leveraging the Alveo board's predefined support files to configure IPs and automate the FPGA constraints mapping. This ensures accurate I/O pinning and streamlines the integration of peripherals into the FPGA fabric. The hardware designs consisting microcontroller unit on the system design are programmed with bare-metal applications developed in Vitis \cite{amd_vitis_ide} to directly access internal memory resources, while PetaLinux tools \cite{PetaLinuxSDK} are used to deploy a lightweight Linux OS on MicroBlaze based configurations \cite{VivadoUG904}. In the Linux environment, low-level utilities such as dd are employed to interact with memory-mapped devices at known addresses defined in the Vivado-generated address map, enabling precise, user-controlled memory access and data handling during experiments.

\subsubsection{BRAM}

On-chip Block RAM (BRAM) is a key feature of FPGA devices, offering low-latency, high-speed memory for temporary data storage and rapid access \cite{fpgamemory}. Unlike off-chip DDR4, BRAM is directly integrated into the FPGA fabric, providing fast access times and deterministic behavior due to its proximity to processing logic, close to  which the BRAM slices reside. However, its limited capacity makes it suitable primarily for applications with smaller data sets or time-critical tasks requiring rapid data processing. In this research, BRAM is utilized and accessed via the XDMA IP core to evaluate its performance.

The simple BRAM-based design utilizes the XDMA IP core to facilitate high-throughput communication between the host system and the on-chip memory over a PCIe interface. The XDMA core is configured to perform scatter-gather DMA operations, ensuring efficient data transfer while minimizing host CPU intervention. Block memory is instantiated using a Block RAM memory generator IP \cite{amd_blk_mem_gen} and a BRAM memory Controller \cite{amd_axi_bram_ctrl}. The controller manages the access to embedded BRAM resources, providing a bridge between a bus interface, i.e, the AXI interface, and the BRAM block. Data is directly routed to and from BRAM via an AXI4 memory-mapped interface. This connection over the AXI interface between XDMA and BRAM eliminates the overhead associated with external memory controllers, as seen with DDR controllers, allowing for streamlined memory transactions optimized for latency-sensitive applications.

Although on-chip memories like BRAM and URAM on the Alveo U250 card offer significantly less storage capacity (2MB BRAM  to 90MB URAM blocks) compared to off-chip DDR4, their low latency, tight coupling with programmable logic, and short on-chip path access times make them ideal for evaluating the efficiency of the PCIe-to-AXI DMA path. This design was included to benchmark the achievable bandwidth when targeting memory resources directly integrated within the FPGA fabric. In this design, we have utilized a small amount of 2MB BRAM available locally in the Super Logic Region (SLR) where our design resides on the FPGA. All our designs in the paper reside in the SLR0 region of the XCU250 FPGA which has its own share of BRAM blocks, DDR4 interface pinouts for their on-chip and off-chip memories.  Evaluating BRAM-based transfers provides good insight into how well the XDMA engine saturates the PCIe link even with minimal memory depth, thereby validating the end-to-end memory interface in simpler FPGA memory hierarchies, utilizing On-chip memories.

\begin{figure}[H]
  \centering
\includegraphics[scale=0.25]{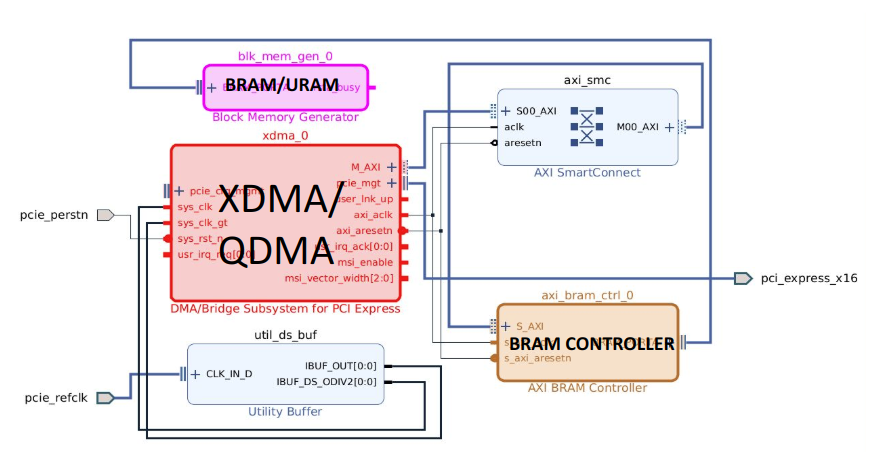} 
\vspace{-5pt}
  \caption{BRAM block diagram design.}
  \label{fig:bram_design2}
\end{figure}

The XDMA IP and its driver leverage H2C (Host-to-Card) and C2H (Card-to-Host) channels to access the BRAM, which is configured to a defined capacity based on the FPGA resources available in the SLR of the Alveo XCU250. Typically, this configuration allows up to 1 MB or 2 MB of BRAM to be accessed, depending on the number of available BRAM blocks. Each channel independently handles data movement in its respective direction, facilitating concurrent operations. The scatter-gather descriptor management by the XDMA driver ensures efficient usage of the BRAM’s limited capacity, with the data divided into smaller chunks for seamless transfer.

By utilizing multiple H2C and C2H channels, the XDMA IP ensures maximum utilization of the PCIe bandwidth while maintaining deterministic access to BRAM. This enables high-speed data transfers suitable for applications such as packet header processing, where low-latency and predictable access times are critical. Figure \ref{fig:bram_design2} illustrates the block diagram of the design using BRAM memory.

\subsubsection{DRAM}

The BRAM's capacity is significantly limited compared to off-chip DDR4 memory, which provides substantially larger storage space. This makes DDR4 a more viable option for applications requiring high-capacity memory, such as large-scale data processing, high-resolution image storage, or network packet buffering in SmartNICs. While BRAM is ideal for low-latency operations or small data sets, DDR4 enables the FPGA to manage workloads with extensive memory requirements, making it a critical component for this research.

The design employs one of the four Alveo cards' DDR4 DIMM memory as the primary off-chip storage, integrated with the XDMA IP core for high-throughput data transfer between the FPGA and host system via PCIe. The 16 GB DDR4 DRAM memory is instantiated using a memory controller,   Memory Interface Generator MIG IP core \cite{amd_memory_ip}, which has the necessary AXI-based interfaces and physical layers to the DRAM memory. The XDMA IP facilitates scatter-gather DMA operations, enabling efficient, large-scale data movement while minimizing host CPU involvement. The DDR4 memory controller, connected through an AXI4 memory-mapped interface, provides a high-speed link to the memory, allowing seamless read and write operations for the FPGA logic. This configuration ensures that data is transferred in alignment with the memory's burst length, optimizing bandwidth utilization and latency.

\begin{figure}[!h]
  \centering
\includegraphics[scale=0.20]{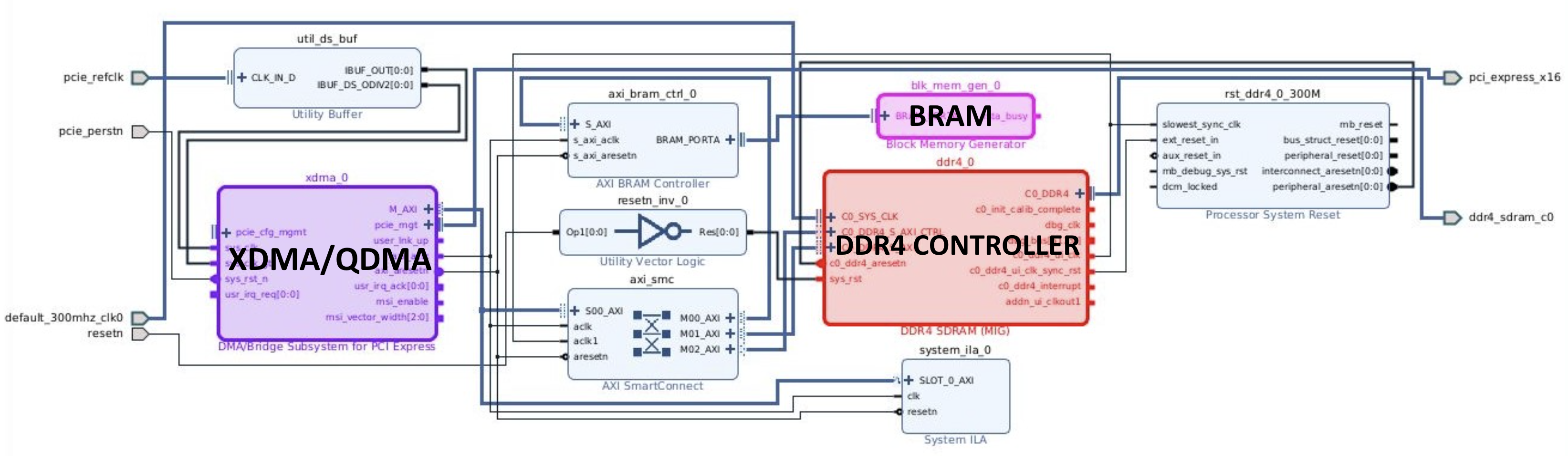} 
\vspace{-5pt}
  \caption{DRAM block diagram design.}
  \label{fig:dram_design}
\end{figure}

In our design, the XDMA core connects to the DDR4 memory subsystem in a streamlined manner using the MIG memory interface controller. This simplicity enables a direct evaluation of DDR4 performance, focusing on maximizing raw data throughput and minimizing latency. Applications for such a design include network packet processing, where large queues of packets are temporarily stored in DDR4, or machine learning inference, where models or intermediate data exceed on-chip memory capacities. This DDR4-based architecture offers a practical solution for bridging the gap between the speed of on-chip memory and the capacity of external storage, providing a balance of high performance and scalability for memory-intensive tasks. Figure \ref{fig:dram_design} illustrates the block diagram of the DRAM-based design.

\subsubsection{DRAM over Microcontroller}
Integrating a MicroBlaze soft processor IP \cite{VivadoUG984} alongside the XDMA IP core in an FPGA-based system enables versatile access to DDR4 memory. This configuration allows DDR4 memory to be accessed both by the host system through the XDMA IP and directly by the MicroBlaze processor for local bare metal processing tasks within the FPGA. Such a setup provides a dual path access mechanism, offering flexibility in managing memory transactions while enabling independent operations by the host and the FPGA fabric.

In this design, the DDR4 controller interfaces with the memory via an AXI4 memory-mapped interface, facilitating high-throughput, low-latency data transfers. The XDMA IP core handles PCIe based data movement between the host and DDR4, leveraging H2C (Host-to-Card) and C2H (Card-to-Host) channels. These channels ensure efficient scatter-gather DMA operations, dividing larger data chunks into smaller, optimized transactions. Simultaneously, the MicroBlaze processor communicates with the DDR4 memory over a separate AXI4 bus, enabling local processing of data directly within the FPGA. This dual access architecture allows the MicroBlaze to perform lightweight computations, such as preprocessing or filtering data, before transferring it to the host system via XDMA. The MicroBlaze soft processor along with the DDR4, was configured to have 32 bit processing capabilities.

This design's flexibility opens up a range of potential applications. For instance, in network acceleration tasks, the MicroBlaze can be used to preprocess packet data stored in DDR4, reducing the computational burden on the host CPU. Similarly, in data acquisition systems, the MicroBlaze can locally process incoming data streams, performing initial aggregation or formatting before transferring the data to the host for further analysis. 

\begin{figure}[!h]
  \centering
\includegraphics[scale=0.15]{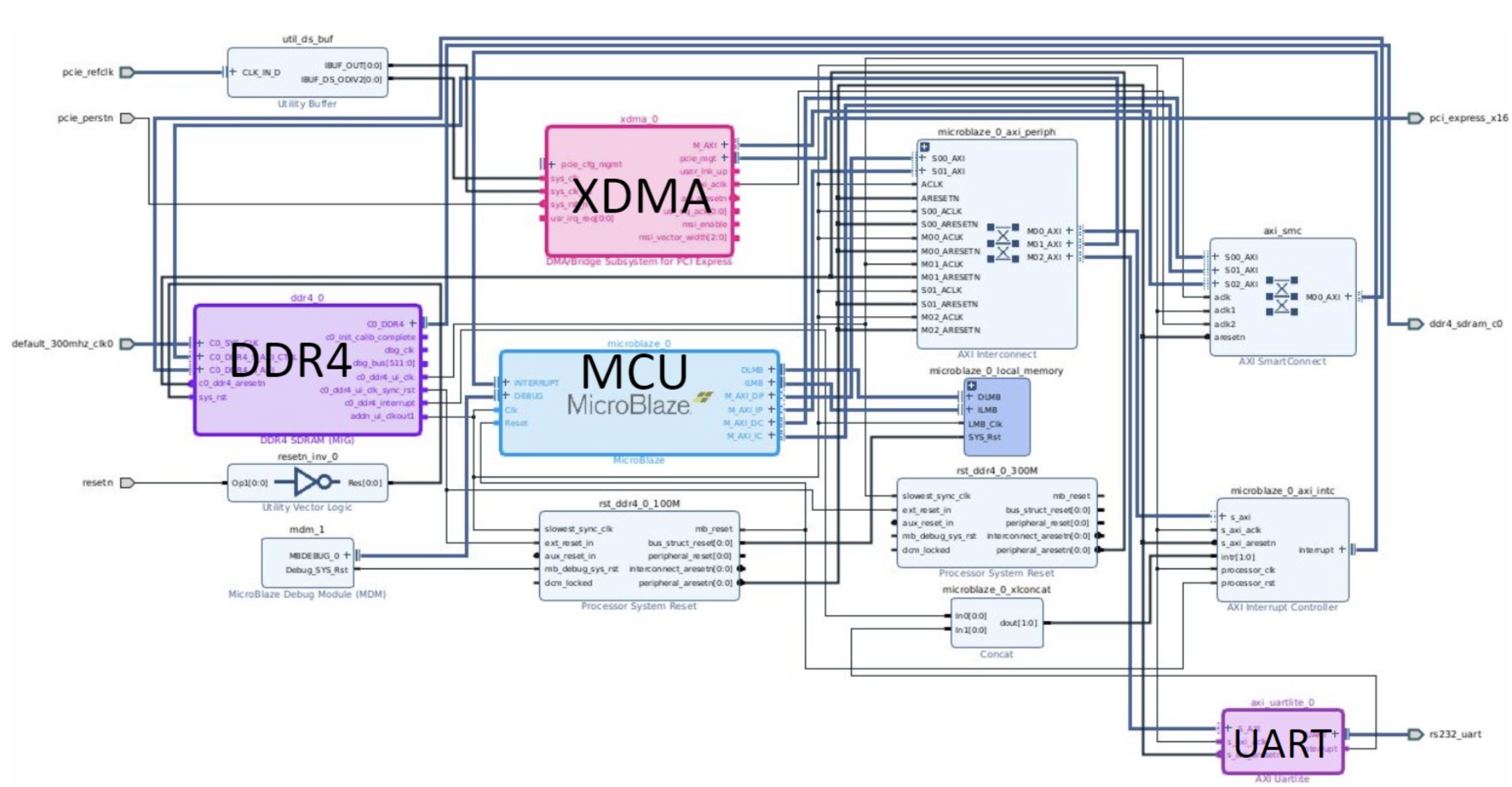} 
\vspace{-15pt}
  \caption{MicroBlaze block diagram design.}
  \label{fig:blaze_design}
\end{figure}

By enabling independent access to DDR4 memory from the host and from within the FPGA, the combination of MicroBlaze and XDMA offers a powerful and scalable solution for memory-intensive FPGA applications, bridging the gap between local FPGA processing and host communication for offloading to smart network devices (i.e., FPGA-based SmartNIC). Figure \ref{fig:blaze_design} depicts the system design featuring MicroBlaze.

\subsubsection{DRAM over PetaLinux}
The integration of a MicroBlaze soft processor running PetaLinux\cite{PetaLinuxSDK} provides a powerful framework for managing DDR4 memory in FPGA-based systems. In this design, the MicroBlaze processor operates as a fully functional embedded processor, booted using a PetaLinux image stored in a Quad SPI (QSPI) flash memory. The QSPI flash is configured to store and initialize the PetaLinux environment using Petalinux tools built on the host system, which facilitates a Linux-based runtime for executing user level programs.

The 1 GB Quad SPI flash memory on the Alveo card is used in the design for bootstrapping the PetaLinux environment. Upon power-up, the FPGA reads the boot image from QSPI, initialize the MicroBlaze processor and load the Linux kernel, drivers, and root filesystem. This enables the FPGA to function autonomously as an embedded system, capable of executing complex tasks independently of the host system.

The DDR4 controller, accessed via an AXI4 memory-mapped interface, provides the MicroBlaze with seamless access to high capacity off chip memory. Additionally, the XDMA IP core enables data transfer between the host system and DDR4 memory over PCIe. The host system utilizes XDMA  Host-to-Card (H2C) and  Card-to-Host (C2H) channels to send data to the FPGA’s DDR4 memory or retrieve results after processing. Concurrently, the MicroBlaze processor running PetaLinux provides a local programming interface to interact with the DDR4 memory, enabling the processor to access data written by the host or perform read-back operations from DDR4.

This setup allows for sophisticated memory management and processing workflows. For instance, the host system can write large data sets to DDR4 using XDMA, and the MicroBlaze can process these data sets locally, applying algorithms for preprocessing, aggregation, or transformation. PetaLinux provides a rich software environment to write and execute programs in user space, enabling efficient interaction with DDR4. Standard Linux APIs, high level languages and drivers developed within PetaLinux streamline memory access, allowing for high-level abstractions when reading or writing to DDR4.

Potential applications for this design include real-time data processing tasks, where the MicroBlaze preprocesses or analyzes data stored in DDR4 before sending results to the host via XDMA. Additionally, it can be used for network packet analysis, where packets sent by the host are buffered in DDR4 and processed locally by the PetaLinux environment. This design is also suited for machine learning workloads, where the MicroBlaze handles intermediate computations or inference tasks while leveraging the DDR4 memory for storage.

\begin{figure}[!h]
  \centering
\includegraphics[scale=0.20]{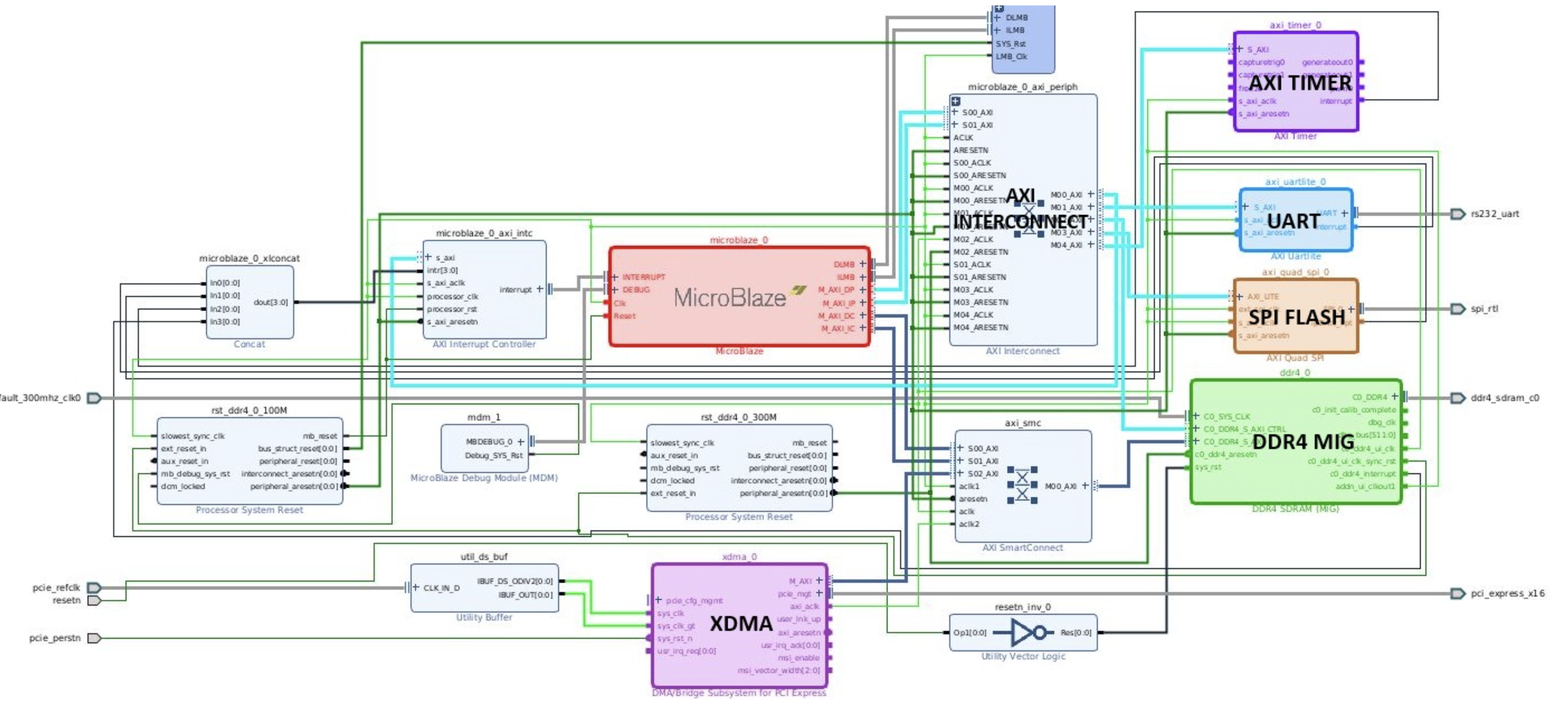} 
\vspace{-15pt}
  \caption{PetaLinux block diagram design.}
  \label{fig:peta_design}
\end{figure}

By combining the flexibility of PetaLinux with the high-speed data transfer capabilities of XDMA over multiple channels, this design gives us a operating system enabled FPGA design that can offload tasks and memory from the host system. Figure \ref{fig:peta_design} illustrates the PetaLinux based design.

\subsection{SOC-based SmartNIC}
In this research, we utilize the NVIDIA BlueField-2 Data Processing Unit (DPU) \cite{bf2}, which supports RDMA and DMA for high-performance communication between host and SmartNIC. It has 8 Arm Cortex-A72 cores, which deliver significant computational power for managing offloaded tasks. It has a single DDR4 DRAM controller, 16GB of on-board DDR4, 8 or 16 lanes of PCIe Gen 4.0 and a dual Ethernet ports of 10/25/50/100 GB/s. 

\subsubsection{RDMA}
The RDMA stack comes on the BlueField-2’s Arm based Linux environment, supporting industry-standard protocols such as RoCE (RDMA over Converged Ethernet) and InfiniBand. These capabilities make BlueField-2 particularly well-suited for data-intensive applications like distributed databases, high-performance computing, and cloud storage, where efficient data transfer and minimal CPU overhead are critical \cite{bayatpour2021bluesmpi}.

\subsubsection{DOCA DMA}

DOCA DMA is provided by NVIDIA as part of the DOCA SDK and is freely available for use. DOCA offers APIs to perform DMA operations on DOCA buffers \cite{dmadoca}. Through these APIs, developers can allocate and register memory buffers (DOCA Buffers) in either the host or the BlueField DPU. These buffers can then be used for zero-copy DMA operations—transferring data directly between devices without intermediate copying steps, which significantly improves performance.

\section{Evaluation}\label{sec:eval}
\begin{figure}[!h]
  \centering
\includegraphics[scale=0.4]{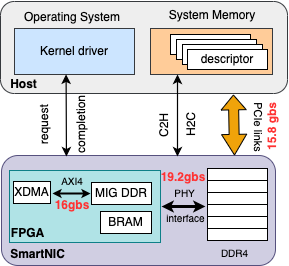} 
\vspace{-3pt}
  \caption{Bandwidth limitations on Alveo 250.}
  \label{fig:model}
\end{figure}

Firstly, we review the bandwidth limitations present along the data path between the host and the SmartNIC. Figure \ref{fig:model} shows the available bandwidth of the Alveo 250 SmartNIC. The PCIe link between the host and the SmartNIC offers a bandwidth of 15.8 GB/s. The physical interface connecting the FPGA to off-chip memory supports 19.2 GB/s, while the AXI4 interconnect within the FPGA provides a bandwidth of 16 GB/s.

There are two DMA operating modes of transfers between the host and the SmartNIC: Host Polled Mode and Host Interrupt Mode. In Host Polled Mode, the CPU actively checks the status of data transfers, resulting in some CPU involvement during DMA operations, which can lead to increased latency. Host Interrupt Mode relies on interrupts from the card (i.e., Alveo card connected via PCIe). This approach reduces CPU involvement, allowing it to perform other tasks until interrupted. While interrupt handling can introduce slightly higher latency, it results in more efficient CPU utilization. The choice between modes depends on application requirements—whether minimizing CPU usage or achieving low-latency transfers is the priority. In our tests, we used MSI-X interrupt mode on the host, as it provided better and more consistent performance.

To validate data access in our designs, we performed memory access by transferring data between the host and the card using the device dump terminal command, i.e., dd \cite{ddman}. This involves issuing raw read and write operations to the addressable memory regions. For evaluating the data transfer performance, we used a combination of dd commands and XDMA driver binaries \cite{XilinxXDMAIPtools}. Data transactions were initiated using the device executable from the XDMA Linux kernel driver suite, which report average bandwidth and timing metrics for transfers over the PCIe H2C and C2H channels. While dd allowed straightforward interaction with device files and quick verification of transactions, the XDMA binaries enabled good control over transfer parameters and repeatability, making them ideal for precise performance characterization. All experiments were run with MSI-X host interrupt mode enabled and optimized PCIe settings, ensuring consistent measurements of bandwidths.

\hfill \newline \hfill

\subsection{XDMA and BRAM performance}

\begin{figure}[!h]
  \centering
\includegraphics[scale=0.35]{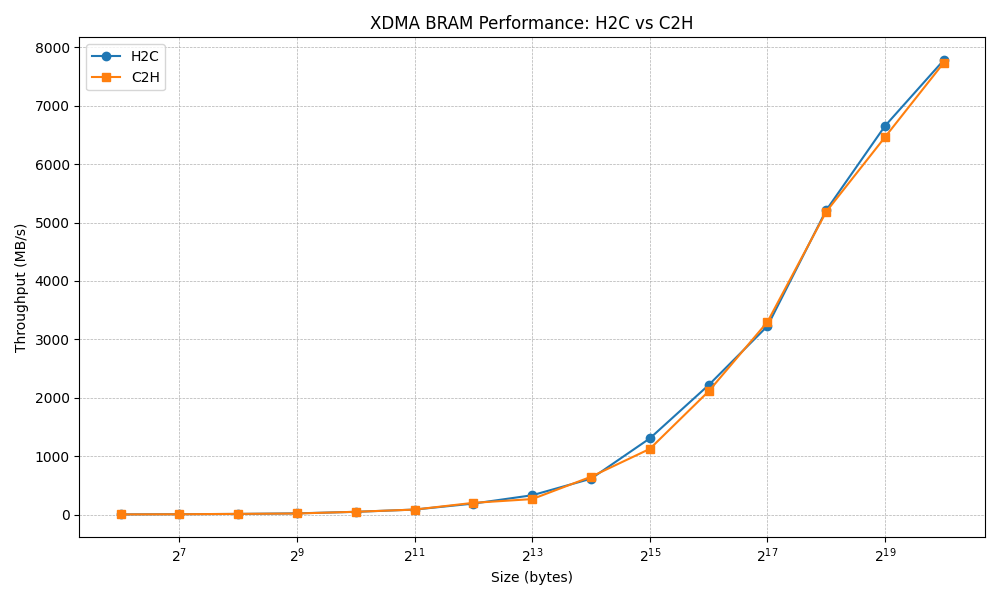} 
\vspace{-10pt}
  \caption{BRAM XDMA single channel performance.}
  \label{fig:bram_design}
\end{figure}

In the XDMA and BRAM design, using a single channel, we observed a consistent rise in bandwidth for both H2C and C2H transfers as the transfer size increased from 64 bytes to 1MB, peaking at 7.54 GB/s for H2C and 7.77 GB/s for C2H as Figure \ref{fig:bram_design} shows. Although limited by the 1 MB of configured  BRAM (out of 8MB available across different slices of the FPGA), the design demonstrated that on-chip memory, despite its smaller size, can still deliver multi-gigabit per second throughput for sub-megabyte transfers. Showing how single-channel configurations can achieve high throughput due to the low latency and direct access paths within the FPGA fabric.

\subsection{XDMA and DDR DRAM performance}

\begin{figure}[!h]
  \centering
\includegraphics[scale=0.35]{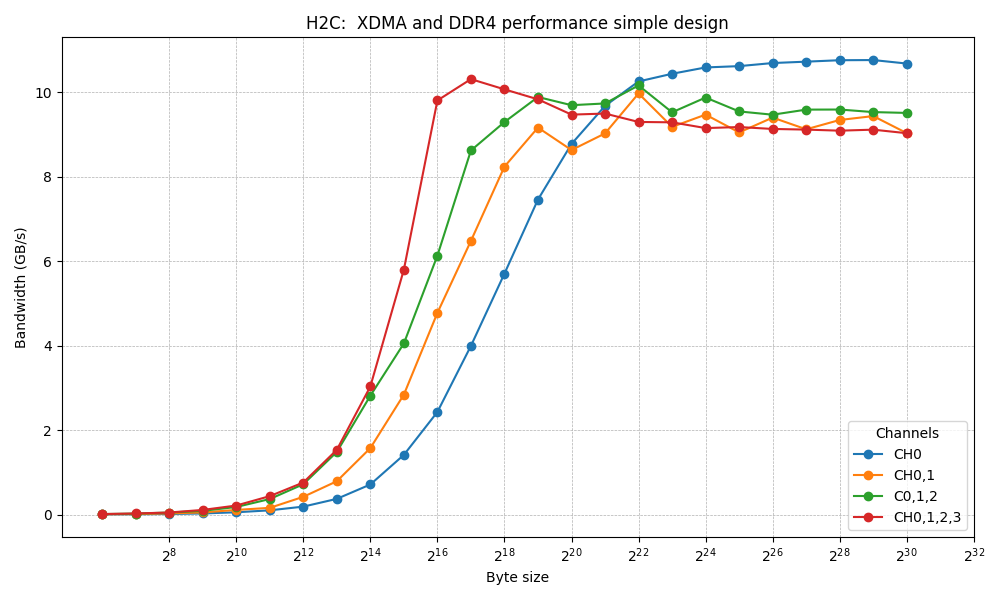} 
\vspace{-10pt}
  \caption{DDR4 Simple design H2C.}
  \label{fig:H2C_simpledesign}
\end{figure}

\begin{figure}[!h]
  \centering
\includegraphics[scale=0.35]{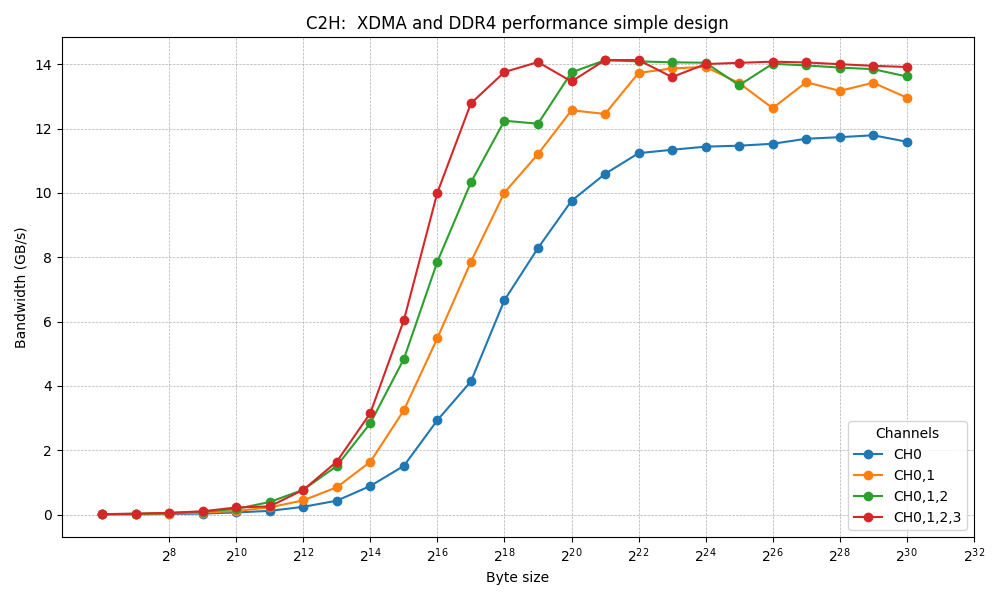} 
\vspace{-10pt}
  \caption{DDR4 Simple design C2H.}
  \label{fig:Simple_C2H}
\end{figure}

In host interrupt mode for XDMA and DDR based simple design, XDMA based H2C and C2H transfers shown in  Figures \ref{fig:H2C_simpledesign} and \ref{fig:Simple_C2H}  exhibit distinct bandwidth scaling and stability characteristics with an increase in the number of XDMA channels in the FPGA. For H2C transfers, bandwidth increases with transfer size, peaking at 10.8 GB/s for a single channel and reaching an aggregated 10 GB/s when multiple interleaved channels are used. Here, we notice mid-range byte sizes exhibit a significant bandwidth boost when using interleaved channels, highlighting the parallelism inherent in the FPGA fabric. FPGAs support multiple channels that can be dynamically managed at runtime based on application demands and system load, enabling concurrent processing of multiple XDMA data streams. This reduces bottlenecks, fully utilizes all PCIe lanes, and maximizes overall throughput. Bandwidth saturation beyond 1 MB occurs at around 9-10 GB/s, with occasional fluctuations. 

Conversely, C2H transfers shown in Fig \ref{fig:Simple_C2H} demonstrate a more stable and higher peak bandwidth, reaching 12 GB/s for a single channel and 13-14 GB/s with interleaved channels, approaching the theoretical 16 GB/s limit of PCIe Gen3 x16. Unlike H2C, C2H exhibits fewer fluctuations, ensuring consistent performance across different transfer sizes. Additionally, host interrupt mode provides greater stability compared to polled mode, making it preferable for high-throughput applications. The results indicate that while both transfer directions benefit from multi-channel configurations, C2H outperforms H2C in terms of peak bandwidth and stability in XDMA DDR4 designs.

\begin{figure}[!h]
  \centering
\includegraphics[scale=0.35]{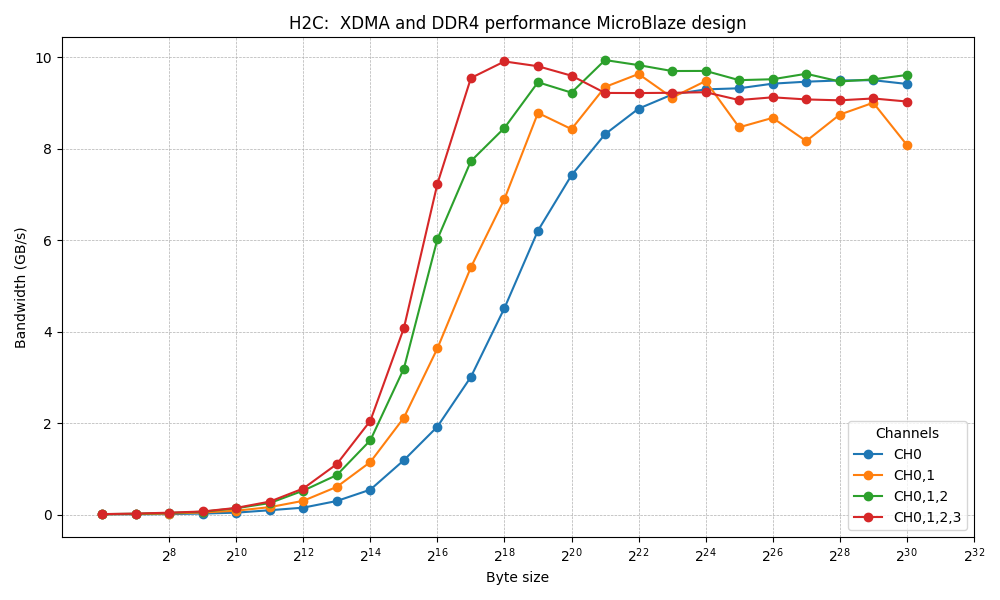} 
\vspace{-10pt}
  \caption{MicroBlaze MCU design H2C.}
  \label{fig:micro_designH2C}
\end{figure}

\begin{figure}[!h]
  \centering
\includegraphics[scale=0.35]{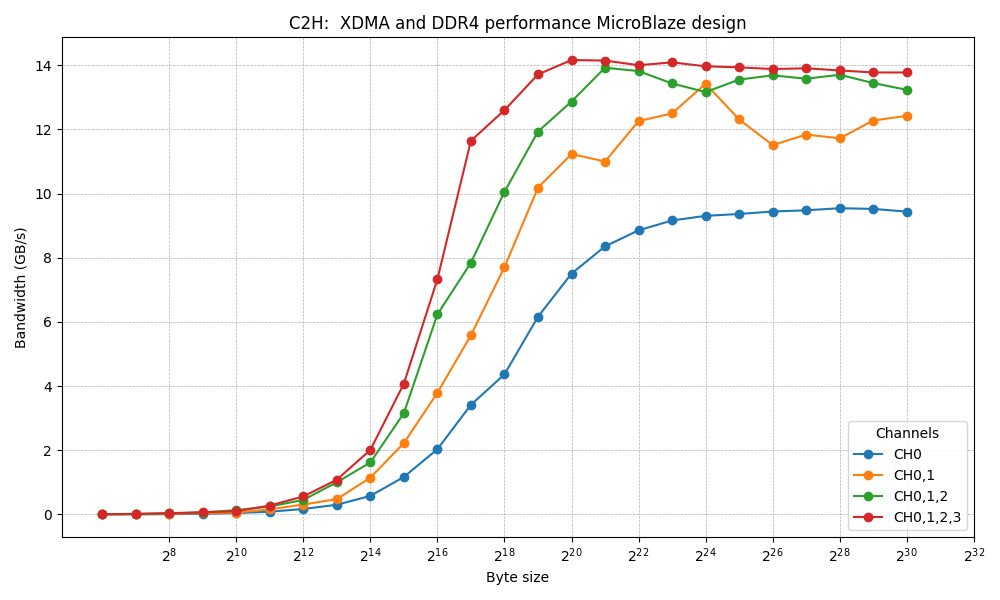} 
\vspace{-10pt}
  \caption{MicroBlaze MCU design C2H.}
  \label{fig:micro_designc2h}
\end{figure}

\subsection{XDMA and DDR performance in MicroBlaze-based design}

The performance of DMA data transfers from host system memory to DRAM, shared between a MicroBlaze CPU Master and the XDMA subsystem master on the FPGA, was evaluated. The MicroBlaze processor was either idle or executing bare-metal memory tests and access routines compiled in Vitis. Testing was conducted with the host set to MSI interrupt mode. Compared to a previous setup featuring only XDMA and DDR, the current design integrates additional peripherals and a new master CPU. From a DMA perspective, a shared DRAM facilitates data exchange between the host system and FPGA components. The H2C results indicate that while the MicroBlaze design in Figure \ref{fig:micro_designH2C} exhibits lower peak bandwidth—around 9.5 GB/s for single H2C channels compared to the 10.7 GB/s rooflining of the simpler setup H2C transfers shown in Figure \ref{fig:H2C_simpledesign}. This reduction in bandwidth can be attributed to contention on the shared AXI interconnect, where both the XDMA engine and the MicroBlaze master compete for access to the DDR4 memory. Even if the MicroBlaze is idle or lightly loaded, the presence of additional peripherals and arbitration logic introduces overhead that slightly degrades the overall throughput achievable by XDMA transfers.

For C2H transfers, an improved bandwidth was observed for aggregated channels, reaching peak values between 13 GB/s and 14 GB/s shown in Figure \ref{fig:micro_designc2h} , consistent with previous observations in C2H transfers with simple DDR XDMA designs. However, single-channel C2H transfers show a drop in bandwidth in the MicroBlaze-based design compared to the simple design without the microcontroller unit. We notice peak bandwidth for a single channel saturates around 9.5 GB/s, lower than the previous 11.8–12 GB/s. Interestingly, both H2C and C2H single-channel transfers maintain nearly identical peak values of 9.4 GB/s, unlike other cases where C2H outperformed H2C. During these experiments, the MicroBlaze processor remained idle, ensuring no contention over the AXI interconnect shared between the XDMA and MicroBlaze masters when accessing DDR memory. To verify this, a loop running NOP instructions on the  MicroBlaze MCU while there is memory tests confirmed that its presence and additional hardware on the shared AXI bus influenced DMA performance.

\subsection{XDMA and DDR performance in Petalinux design}
\begin{figure}[!h]
  \centering
\includegraphics[scale=0.35]{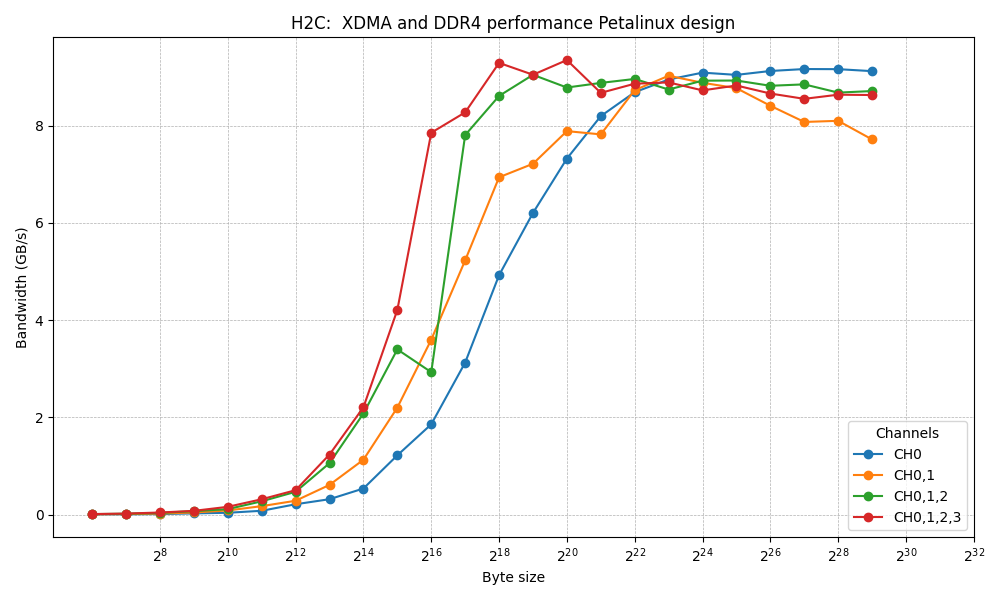} 
\vspace{-10pt}
  \caption{Petainux H2C.}
  \label{fig:peta_designh2c}
\end{figure}

\begin{figure}[!h]
  \centering
\includegraphics[scale=0.35]{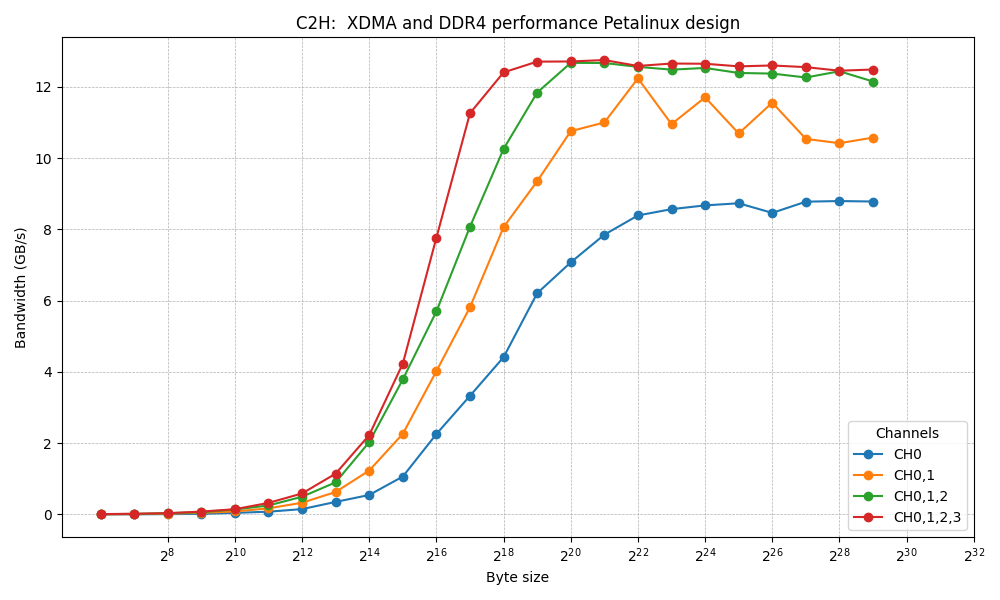} 
\vspace{-10pt}
  \caption{Petalinux C2H.}
  \label{fig:peta_designc2h}
\end{figure}

When we have a Petalinux kernel running on the MicroBlaze design with additional peripherals, we write to that part of the DDR4 memory space of the FPGA card where the kernel doesn't reside, to make sure that the operating system doesn't crash unwillingly. The memory space is shared between the XDMA master and MicroBlaze, which has kernel operations and memory management running actively, hence an AXI bus contention between the masters and a overhead is expected to affect the DMA performance.  

For H2C transfers shown in Fig \ref{fig:peta_designh2c}, we achieve a maximum bandwidth peaking at 9.2 GB/s for single H2C channel transfers. Interleaved channels give us better performance for most byte sizes until the saturated max bandwidth of 9 GB/s. These numbers do not fall much behind the design which had just a bare metal MicroBlaze peripheral in it, showing that the FPGA’s internal AXI interconnects enable efficient parallel DMA transfers, despite the overhead caused by the kernel operations from the MicroBlaze master. 

While the H2C transfers were not all stable for multiple channels, we see more stable performance in C2H transfers as seen in the Fig \ref{fig:peta_designc2h}. We reach good bandwidths, roof-lining at 12-13 GB/s, three and four C2H channel transfers. Two-channel transfers are seen fluctuating in almost every design iteration including this, achieving a two-channel bandwidth between 10.5-12 GB/s.

\begin{figure}[!h]
  \centering
\includegraphics[scale=0.42]{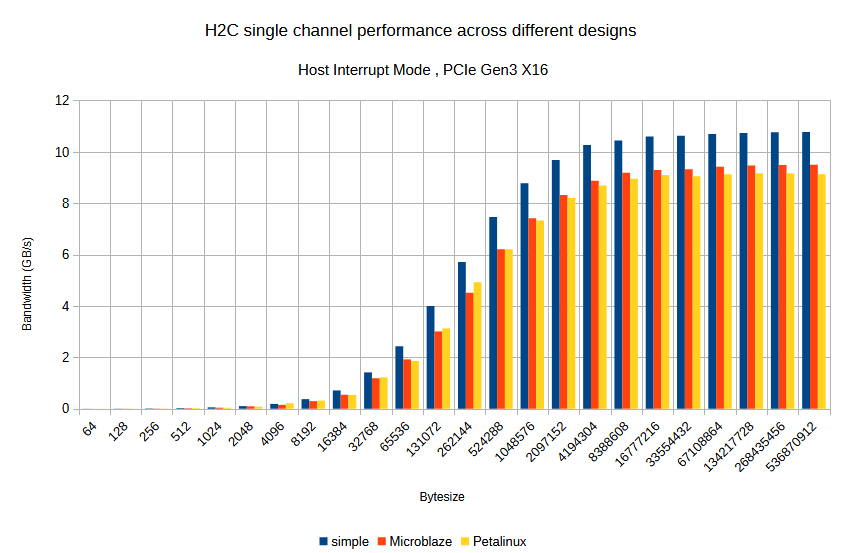} 
\vspace{-10pt}
  \caption{Single Channel H2C.}
  \label{fig:single_H2X}
\end{figure}

\begin{figure}[!h]
  \centering
\includegraphics[scale=0.42]{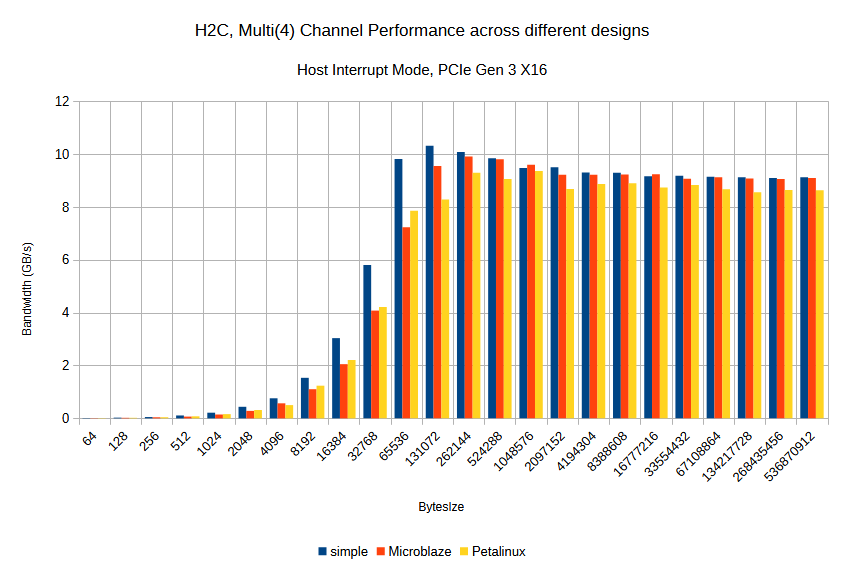} 
\vspace{-10pt}
  \caption{Multi-Channel H2C.}
  \label{fig:MultiH2C}
\end{figure}

The comparison between single-channel and multi-channel H2C XDMA performance shown in figures \ref{fig:single_H2X} and \ref{fig:MultiH2C} highlights the advantages of utilizing multiple DMA engines for data transfers over the PCIe link. In the single-channel host-to-card transfer case seen in Figure \ref{fig:single_H2X}, performance is limited by the bandwidth of a single DMA channel, which gradually increases with larger byte sizes and peaks at approximately 9–10 GB/s. However, when XDMA is configured with four independent channels, the available PCIe bandwidth is utilized more efficiently, leading to higher sustained bandwidth and a steeper rise in bandwidths for mid range bytesizes, which later saturate at higher sizes. This demonstrates the ability of FPGA-based multi-channel XDMA to effectively parallelize data transfers, reducing contention and improving overall bandwidth utilization.

Additionally, the results indicate that different FPGA-based designs, including one running MicroBlaze and another operating with PetaLinux, can achieve comparable performance at larger byte sizes. Despite the added overhead of an embedded Linux environment, the PetaLinux based design still reaches near-maximum throughput, showing that a well-optimized software stack can fully leverage the multi-channel DMA capabilities of the FPGA. This is particularly beneficial for data-intensive applications such as AI inference, high-speed networking, and real-time processing, where maximizing PCIe bandwidth is crucial.

\begin{figure}[H]
  \centering
\includegraphics[scale=0.42]{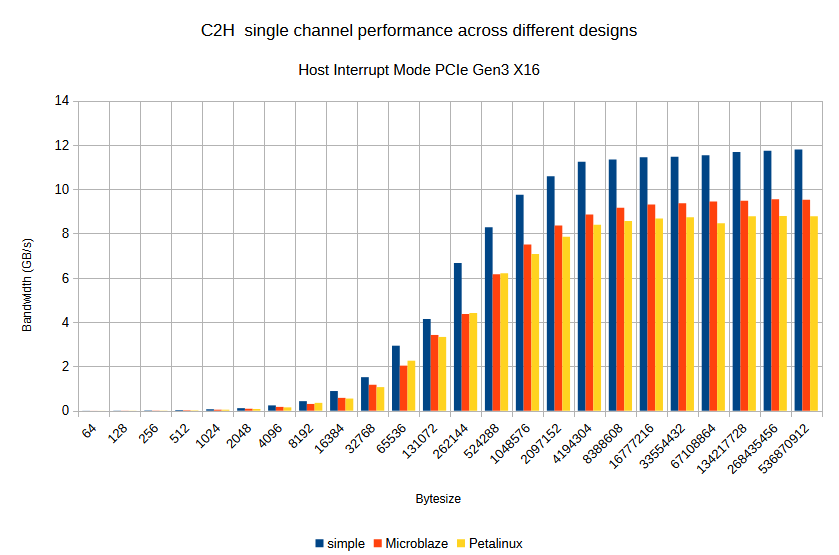} 
\vspace{-10pt}
  \caption{Single Channel C2H.}
  \label{fig:single_C2H}
\end{figure}

\begin{figure}[H]
  \centering
\includegraphics[scale=0.42]{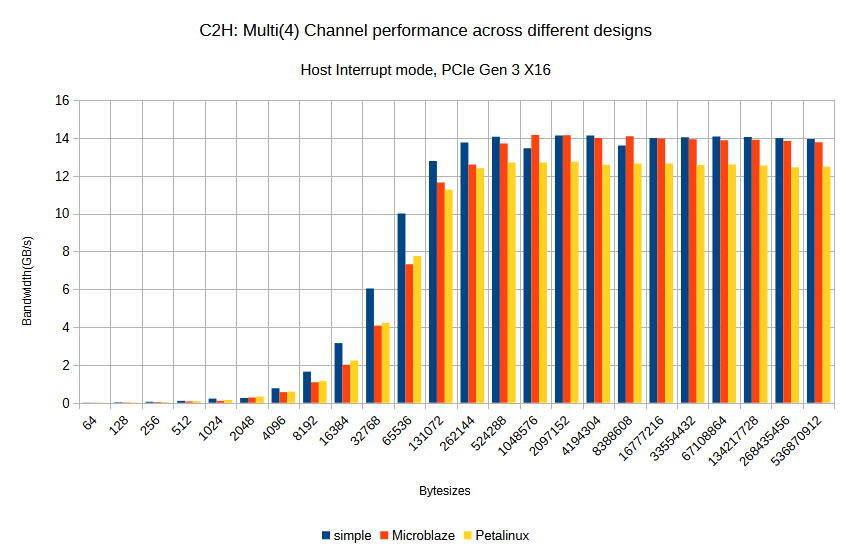} 
\vspace{-10pt}
  \caption{Multi-Channel C2H.}
  \label{fig:Multi_C2H}
\end{figure}

The comparison of C2H  transfers across different FPGA-based designs further reinforces the benefits of utilizing multi-channel XDMA for maximizing PCIe bandwidth. In the single-channel case, Figure \ref{fig:single_C2H}, performance gradually increases with larger byte sizes, peaking at approximately 12–13 GB/s. However, when employing four XDMA channels, the bandwidth utilization improves significantly, allowing the system to sustain higher transfer rates across a wider range of byte sizes, improving C2H bandwidths even for complex designs.

Where increased complexity and system overhead in the FPGA design would be expected to reduce bandwidths at higher transfer sizes, the multi-channel C2H performance remains highly competitive even for larger byte sizes seen in Figure \ref{fig:Multi_C2H}. The ability to sustain peak bandwidth even at high byte sizes showcases the efficiency of the XDMA implementation, ensuring that PCIe resources are fully utilized.

\subsection{RDMA Performance in SoC}
We evaluated RDMA performance for both memory reads and writes using the widely adopted perftest benchmarking suite, which is specifically designed to measure low-level RDMA throughput and latency \cite{perftest}. In particular, we ran the ib\_read\_bw and ib\_write\_bw tests across various message sizes. 

\begin{figure}[!h]
  \centering
\includegraphics[scale=0.35]{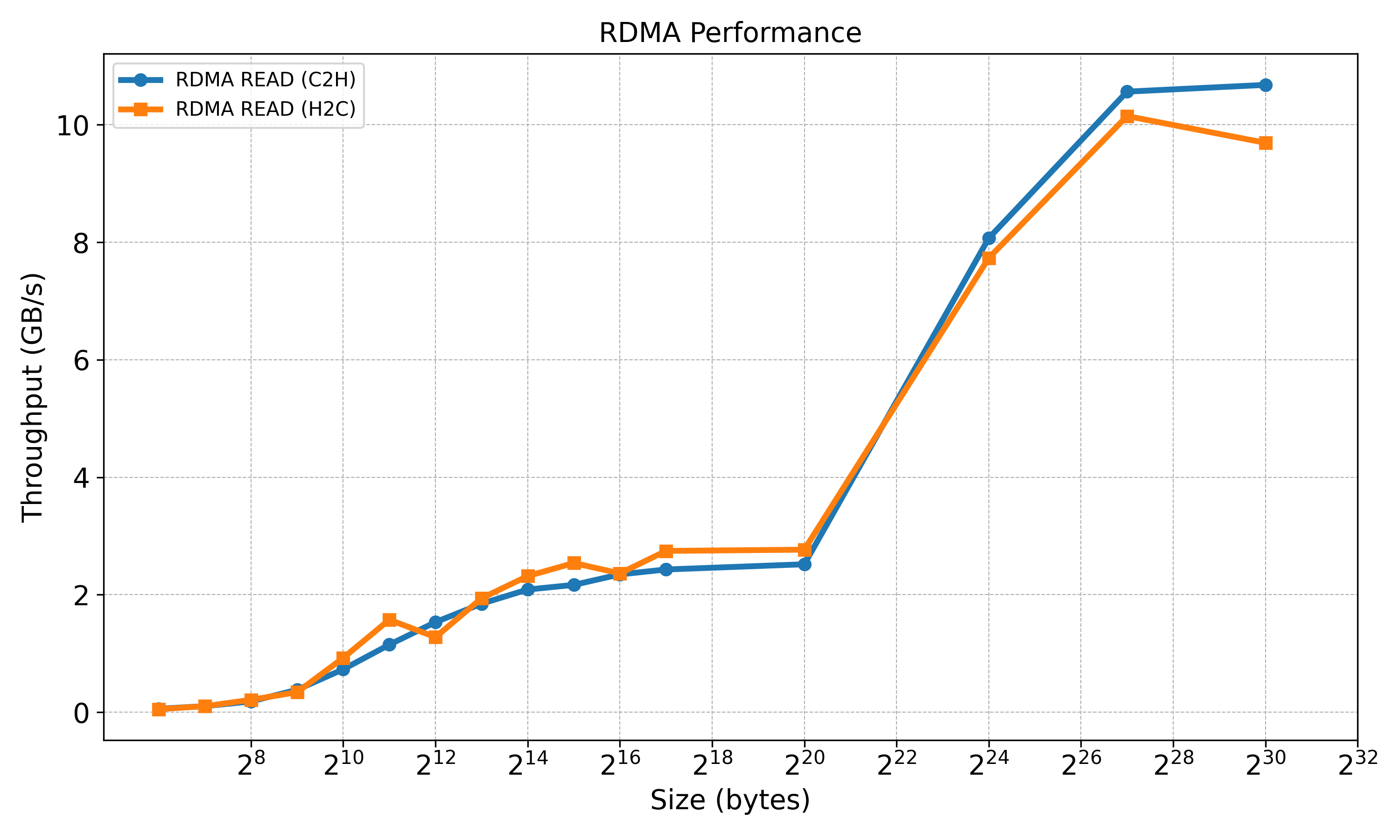} 
\vspace{-10pt}
  \caption{RDMA READ Performance.}
  \label{fig:rdma_read_design}
\end{figure}

\begin{figure}[!h]
  \centering
\includegraphics[scale=0.35]{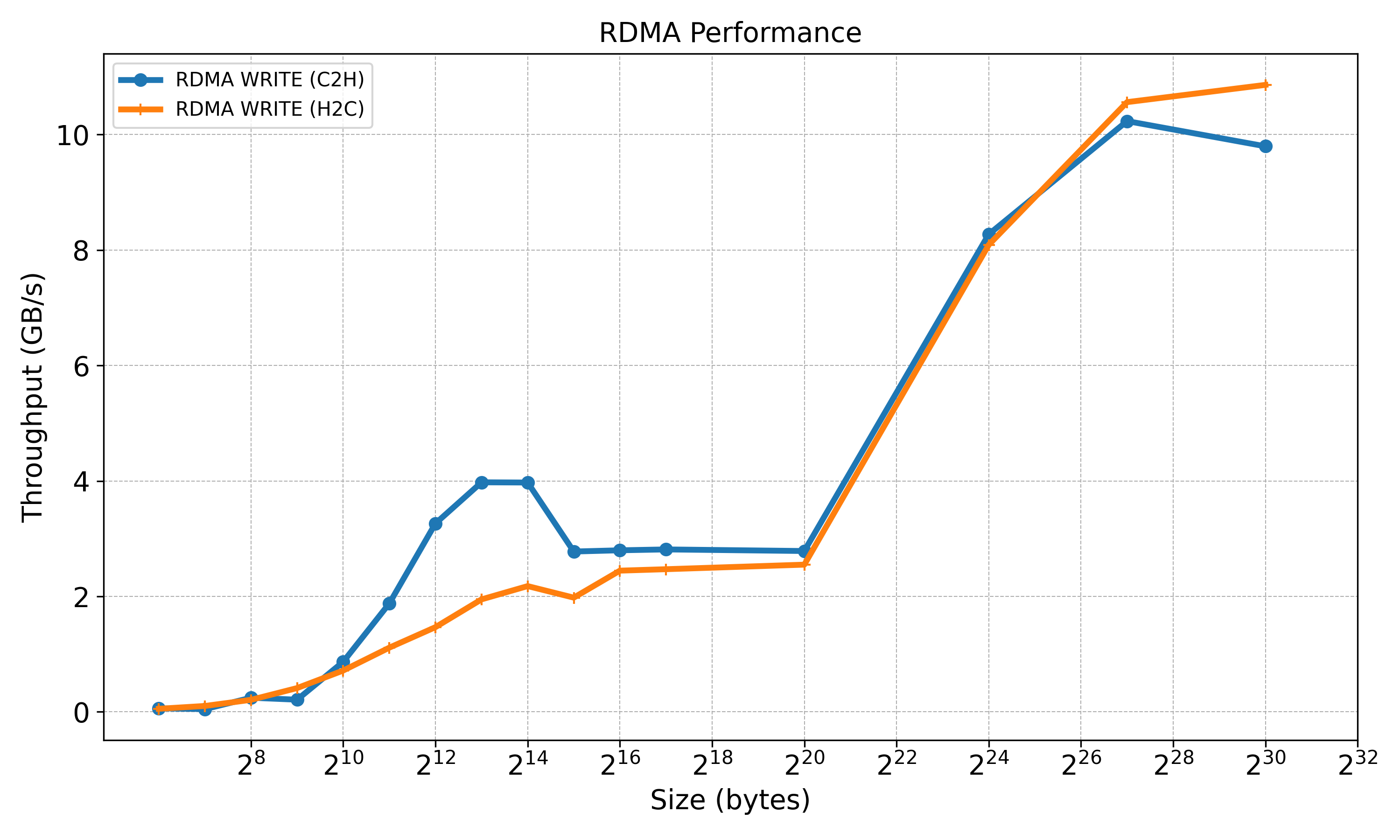} 
\vspace{-10pt}
  \caption{RDMA WRITE Performance.}
  \label{fig:rdma_write_design}
\end{figure}

The results of these experiments are presented in Figures \ref{fig:rdma_read_design} and \ref{fig:rdma_write_design}, which compare performance across two communication directions. In the C2H configuration, the RDMA server is executed on the host system, while the client runs on the SmartNIC. This setup allows the SmartNIC to initiate memory operations targeting the host memory. On the other hand, the H2C scenario represents the reverse configuration, where the server runs on the SmartNIC and the client on the host, thereby enabling the host to access memory located on the SmartNIC. As observed, RDMA memory access performance is lower than that of FPGA-based and custom-designed solutions, especially those operating on bare metal. However, FPGA programming is significantly more complex than programming for RDMA memory access.

\section{Conclusion and future work}\label{sec:conc}

This study presents a detailed evaluation of host to SmartNIC memory access, and vice versa, with a specific focus on leveraging XDMA over PCIe to access different memory configurations, including BRAM and DDR4. Our experiments reveal that lightweight designs utilizing on-chip BRAM can deliver high throughput for sub-megabyte transfers, while DDR4-based configurations, especially with multi-channel XDMA, maximize PCIe bandwidth utilization. We further show that more complex designs involving MicroBlaze and PetaLinux, though subject to AXI interconnect contention, still maintain comparable bandwidth at higher transfer sizes with interleaved DMA channels designed on the FPGA, demonstrating the efficiency and scalability of the FPGA fabric and XDMA architecture.

This foundational work lays the groundwork for future comparative studies across different SmartNIC platforms. As a next step, we plan to expand our benchmarking to include SOC-based SmartNICs such as NVIDIA's BlueField DPU using DOCA DMA and RDMA. By comparing FPGA-based and SoC-based memory access mechanisms under real-world workloads, we aim to develop deeper insights into the performance trade-offs of programmable network devices in in-network computing.

We intend to broaden our evaluation by analyzing memory access under various criteria, such as message rate and diverse access patterns. In addition, we plan to implement memory-mapped I/O to gain deeper insights into performance and efficiency. We have initiated preliminary testing with DMA DOCA; however, an in-depth analysis is planned for future work. Additionally, we plan to evaluate the DOCA GPUNetIO Library, which facilitates communication between GPU memory and the DPU \cite{DOCAGPUNetIO}. Furthermore, we aim to assess memory performance in real-world applications and analyze the impact of memory access on their overall performance.

\section*{Acknowledgment}
The research presented in this paper has benefited from the Experimental Infrastructure for Exploration of Exascale Computing (eX3), which is financially supported by the Research Council of Norway under contract 270053.

\bibliography{ref}

\end{document}